\documentstyle[aps,preprint]{revtex}
\begin{document}
% \draft command makes pacs numbers print
\draft
% repeat the \author\address pair as needed
\title{\bf Quantum state reconstruction in the presence of dissipation}
\author{H. Moya-Cessa\footnote{Permanent address: INAOE, Coordinaci\'on de Optica, 
Apdo. Postal 51 y 216, 72000 Puebla, Pue., Mexico.}\footnote{Electronic address: hmmc@inaoep.mx}, 
S.M. Dutra\footnote{Electronic address: dutra@rulhm1.leidenuniv.nl}, 
J.A. Roversi\footnote{Electronic address: roversi@ifi.unicamp.br}, 
and A. Vidiella-Barranco\footnote{Electronic address: vidiella@ifi.unicamp.br} 
\footnote{Phone: +55 019 7885442; FAX: +55 019 7885427}}
\address{Instituto de F\'\i sica ``Gleb Wataghin'',
Universidade Estadual de Campinas,
13083-970   Campinas  SP  Brazil}
\date{\today}
\maketitle
\begin{abstract}
We propose a realistic scheme to determine the quantum state of a single mode
cavity field even after it has started to decay due to the coupling with an 
environment.
Although dissipation destroys quantum coherences, we show that at zero 
temperature enough information about the initial state remains, in an 
observable quantity, to allow the reconstruction of its Wigner function.
\end{abstract}
% insert suggested PACS numbers in braces on next line
\pacs{42.50.-p, 03.65.Bz, 42.50.Dv}

Methods to reconstruct quantum states of light are of great importance
in quantum optics. There have been several proposals using different 
techniques to achieve such reconstructions \cite {ul}, amongst them, 
the direct sampling of the density matrix of a signal mode in optical 
homodyne tomography \cite {1}, the tomographic reconstruction by unbalanced 
homodyning \cite {2}; the direct measurement (quantum endoscopy), of the 
Wigner function of the electromagnetic field in a cavity or the vibrational 
state of an ion in a trap \cite {3,4}. It is well known that
dissipation has a destructive effect in most of these schemes, and issues
such as compensation of losses in quantum-state measuments have already been 
discussed in the literature \cite{u}. 

In this contribution we present a novel method of how to reconstruct a 
quantum state even after the action of dissipation. 
We consider a single mode high-$Q$ cavity where a
nonclassical field state $\hat{\rho}(0)$ is prepared and subsequently driven
by a coherent pulse. Both processes are assumed to occur in a time scale
much shorter than the decay time of the cavity. Then the field is allowed 
to decay. We will show below that by displacing the initial state we make
its quantum coherences robust enough to allow its experimental 
determination despite the existence of dissipation. 

The master equation in the interaction picture for a damped cavity mode
at zero temperature and under the Born-Markov approximation is given by
\cite {5}

\begin{eqnarray}
         \frac{ \partial \hat{\rho}}{\partial t} = \frac{\gamma}{2} \left(
           2\hat{a} \hat{\rho} \hat{a}^\dagger 
             - \hat{a}^\dagger \hat{a} \hat{\rho} - \hat{\rho} \hat{a}^\dagger
             \hat{a}\right),    \label{1}
\end{eqnarray} 
where $\hat{a}$ and $\hat{a}^\dagger$ are the annihilation and
creation operators and $\gamma$ the decay constant. We can define the
superoperators $\hat{J}$ and $\hat{L}$ by their action on the density
operator \cite {6}

\begin{equation}
\hat{J}\hat{\rho} = \gamma \hat{a}\hat{\rho}\hat{a}^\dagger, \ \ \ \ 
\hat{L}\hat{\rho} = -\frac{\gamma}{2} \left(\hat{a}^\dagger \hat{a} \hat{\rho} +
\hat{\rho} 
\hat{a}^\dagger\hat{a}\right).    \label{3}
\end{equation}

The formal solution of (\ref{1}) can be written as \cite {7}

\begin{equation}
\hat{\rho}(t) = \exp\left[(\hat{J}+\hat{L})t\right]\hat{\rho}(0) 
        =\exp(\hat{L}t)\exp\left[\frac{\hat{J}}{\gamma}
          (1-e^{-\gamma t})\right]\hat{\rho}(0). \label{4}
\end{equation}

We assume that the initial field $\hat{\rho}(0)$ is prepared in a time scale much
shorter than the decay time of the cavity $\gamma^{-1}$. As soon as the field
is generated, a coherent field $|\alpha\rangle$ is injected inside the cavity 
(also in a short time scale) displacing then the initial state \break
$\hat{\rho}_\alpha=\hat{D}(\alpha)\hat{\rho}(0)\hat{D}^{\dagger}(\alpha)$. 
This procedure will enable us to
obtain information about all the elements of the initial density
matrix from the diagonal elements of the time-evolved displaced density matrix
only. As diagonal elements decay much slower than off-diagonal ones, 
information about the initial state stored this way becomes robust enough to 
withstand the decoherence process. We will now show how this robustness can be
used to obtain the Wigner function of the initial state after it has started to
decay.

The diagonal matrix elements of 
$\hat{\rho}_\alpha (t)=\exp\left[(\hat{J}+\hat{L})t\right]\hat{\rho}_\alpha$ 
in the number state basis are 

\begin{equation}
\langle m|\hat{\rho}_\alpha (t)|m\rangle=\frac{e^{-m\gamma t}}{q^m}\sum_{n=0}^{\infty}
q^n \left(\begin{array}{c} n \\ m \end{array} \right)\langle n|
\hat{\rho}_\alpha |n\rangle,
\end{equation}
where $q=1-e^{-\gamma t}$.

We note that if we multiply those elements by the function 

\begin{equation}
\chi(t)=1-2e^{\gamma t} \label{chi}
\end{equation}
and sum over $m$ we obtain

\begin{equation}
F=\frac{2}{\pi}\sum_{m=0}^\infty \chi^m(t)\langle m|\hat{\rho}_\alpha (t)|m\rangle=
\frac{2}{\pi}\sum_{n=0}^\infty (-1)^n\langle n|\hat{D}(\alpha)\hat{\rho}(0)
\hat{D}^{\dagger}(\alpha)|n\rangle.\label{trans}
\end{equation}

The expression above is exactly the Wigner function corresponding to 
$\hat{\rho}$ (the initial field state) \cite{8} at the point specified by the
complex amplitude $\alpha$. Therefore if we measure the diagonal elements 
of the dissipated displaced cavity field 
$P_m(\alpha;t)=\langle m|\hat{\rho}_\alpha (t)|m\rangle$ for a range of $\alpha$'s, the 
transformation in Eq. (\ref{trans}) will give us the Wigner function $F$
for this range. This is the main result of our paper; the reconstruction is made
possible even under the normally destructive action of dissipation. We would like to
stress that the identity in Eq. (\ref{trans}) means that the time-dependence is completely
cancelled, bringing out the Wigner function of the initial state.

One way of determining $P_m(\alpha;t)$ is by injecting 
atoms into the cavity and measuring their population inversion as they exit
after an interaction time $\tau$ much shorter than the cavity decay time.
We may use three-level atoms in a cascade configuration with the upper and the
lower level having the same parity. In this case the population inversion 
is given by \cite{dan}
\begin{equation}
W(\alpha;t+\tau)=\sum_{n=0}^{\infty}P_n(\alpha;t)\left[\frac{\Gamma_n}{\delta^2_n}+
\frac{(n+1)(n+2)}{\delta^2_n}
\cos\left(2\delta_n\lambda\tau\right)\right],\label{inv}
\end{equation}
where $\Gamma_n=\left[\Delta+\chi(n+1)\right]/2$, $\delta_n^2=\Gamma_n^2+\lambda^2(n+1)(n+2)$,
$\Delta$ is the atom-field detuning, $\chi$ is the Stark shift coefficient, and 
$\lambda$ is the coupling constant. In the case of having $\Delta=0$ (two-photon resonance 
condition), $\chi=0$, and for strong enough fields, for which it is valid the approximation
$\left[(n+1)(n+2)\right]^{1/2}\approx n+3/2$, the population inversion reduces to
\begin{equation}
W(\alpha;t+\tau)=\sum_{n=0}^{\infty}P_n(\alpha;t)
\cos\left(\left[2n+3\right]\lambda\tau\right),\label{invs}
\end{equation}  

By inverting the Fourier series in Eq. (\ref{invs}) we obtain for 
$P_n(\alpha;t)$
\begin{equation}
P_n(\alpha;t)=\frac{2\lambda}{\pi}\int_0^{\frac{\pi}{\lambda}} d\tau\:W(t+\tau)
\cos\left(\left[2n+3\right]\lambda\tau\right). \label{pn}
\end{equation}
We need a maximum interaction time $\tau_{max}=\pi/\lambda$  
much shorter than the cavity decay time. This condition implies that we must
be in the strong-coupling regime, i.e. $\lambda\gg\gamma$.

Our scheme is easily generalized to other ($s$-parametrized \cite{9}) 
quasi-probability distributions given by \cite{8}, 
\begin{equation}
F(\alpha;s)=-\frac{2}{\pi(s-1)}\sum_{n=0}^\infty \left(\frac{s+1}{s-1}\right)^n
\langle n|\hat{\rho}_\alpha |n\rangle,
\end{equation}
by choosing
\begin{equation}
\chi(s;t)=1+\frac{2e^{\gamma t}}{s-1}.
\end{equation}

In conclusion, we have presented a novel technique to reconstruct 
the Wigner function of an initial nonclassical state at times
when the field would have normally lost its quantum coherence \cite{nwig}.
Reconstruction approaches do not usually take into account the effect
of losses. The crucial point of our method is the driving of the initial
field immediately after preparation, that is not only used to
cover a region in phase space but also to store quantum coherences
in the diagonal elements of the time evolved displaced density
matrix, making them robust. In other words, we have shown that the
initial displacement transfers to any initial state
the robustness of a coherent state \cite{zurek} against dissipation.

The possibility of reconstructing quantum states at any time opens
up potential applications in quantum computing. For instance, this
method could be used in a scheme to refresh the state of a quantum computer
in order to avoid dissipation-induced errors.

\acknowledgements

One of us, H.M.-C., thanks W. Vogel for useful comments.
This work was partially supported by  Funda\c c\~ao de Amparo \`a
Pesquisa do Estado de S\~ao Paulo (FAPESP), Brazil, Consejo Nacional de
Ciencia y Tecnolog\'\i a (CONACyT), M\'exico, Conselho Nacional de 
Desenvolvimento Cient\'\i fico e Tecnol\'ogico (CNPq), Brazil, and 
International Centre for Theoretical Physics (ICTP), Italy.


\begin{references}

\bibitem{ul} 
{\sc Leonhardt, U.}, 1997, {\em Measuring the Quantum State of Light} 
(CUP, Cambridge), and references therein.
\bibitem{1} 
{\sc Zucchetti, A., Vogel, W., Tasche, M.}, and {\sc Welsch, D.-G.}, 1996, 
{\it Phys. Rev.} A, {\bf 54},1678.
\bibitem{2} 
{\sc Wallentowitz, S.} and {\sc Vogel, W.}, 1996, {\it Phys. Rev.} A, {\bf 53}, 4528;
{\sc Banaszek, K.} and {\sc W\'odkiewicz, K.}, 1996, {\it Phys. Rev. Lett.}, {\bf 76}, 
4344.
\bibitem{3} 
{\sc Lutterbach, L.G.} and {\sc Davidovich, L.}, 1997, {\it Phys. Rev. Lett.}, {\bf 78}, 
2547. 
\bibitem{4} 
{\sc Bardroff, P.J., Leichtle, C., Schrade, G.}, and {\sc Schleich, W.P.}, 1996, 
{\it Phys. Rev. Lett.}, {\bf 77}, 2198.
\bibitem{u} {\sc Kiss, T., Herzog, U.}, and {\sc Leonhardt, U.}, 1995, {\it Phys. Rev.} A, 
{\bf 52}, 2433; {\sc D'Ariano, G.M.D.}, and {\sc Macchiavello, C.}, 1998, {\it Phys. Rev.} A, 
{\bf 57}, 3131.
\bibitem{5}
{\sc Louisell, W.H.}, 1973, {\em Quantum Statistical Properties of Radiation}
(Wiley, New York).
\bibitem{6}
{\sc Barnett, S.M.,} and {\sc Knight, P.L.}, 1986, {\it Phys. Rev.} A, {\bf 33}, 2444.
\bibitem{7}
{\sc Barnett, S.M.}, 1985, Ph.D. thesis (University of London);
{\sc Phoenix, S.J.D.}, 1990, {\it Phys. Rev.} A, {\bf 41}, 5132.
\bibitem{dan}
{\sc Moya-Cessa, H.}, {\sc Knight, P.L.}, and {\sc Rosenhouse-Dantsker, A.} 1994, 
{\it Phys. Rev.} A, {\bf 50}, 1814.
\bibitem{9} 
{\sc Cahill, K.E.} and {\sc Glauber, R.J.}, 1969, {\it Phys. Rev.}, {\bf 177}, 1882.
\bibitem{8}
{\sc Moya-Cessa, H.}, and {\sc Knight, P.L.}, 1993, {\it Phys. Rev.} A, {\bf 48}, 2479.
\bibitem{nwig} At such times the Wigner function would have normally lost its negativity
reflecting, the loss of quantum coherences.
\bibitem{zurek} {\sc Zurek, W.H., Habib  S.}, and {\sc Paz J.P.}, 1993, {\it Phys. Rev. 
Lett.} {\bf 70}, 1187; {\sc Dutra S.M.}, 1998, {\it J. mod. Optics}, {\bf 45}, 759.

\end{references}
\end{document}